\documentclass[prl,aps,floats,twocolumn]{revtex4}

\usepackage[dvips]{graphicx}
\usepackage{amssymb} 
\usepackage{amsmath}

\begin{document}

\title{Testing Inflation: A Bootstrap Approach}

\author{Latham Boyle$^{1}$ and Paul J. Steinhardt$^{2,3}$}

\affiliation{$^1$Canadian Institute for Theoretical Astrophysics
  (CITA), Toronto, Canada \\
  $^2$Princeton Center for Theoretical Science, Princeton University,
  Princeton, NJ 08544 \\
  $^3$Department of Physics, Princeton University, Princeton, NJ
  08544}

\date{October 2008}
 
\begin{abstract}
  We note that the essential idea of inflation, that the universe
  underwent a brief period of accelerated expansion followed by a long
  period of decelerated expansion, can be encapsulated in a ``closure
  condition'' which relates the amount of accelerated expansion during
  inflation to the amount of decelerated expansion afterward.  We
  present a protocol for systematically testing the validity of this
  condition observationally.
\end{abstract}
\maketitle 

Is it possible to show convincingly that inflation \cite{inflation,
  Weinberg} is responsible for the large-scale homogeneity, isotropy
and flatness of the universe, and the primordial spectrum of metric
fluctuations that seeded galaxy formation and sourced the temperature
and polarization variations in the cosmic microwave background (CMB)?
Some would claim no, because there is too much freedom in constructing
inflationary models.  Even if one shows that the observations are
consistent with the predictions of a particular inflationary model,
this is unconvincing because, for virtually any given combination of
observations, one can design many inflationary models that reproduce
them.  If a theory allows everything, it has no predictive power.

In this paper, though, we describe how to combine observations into a
sequence of ``bootstrap tests'' that, if any one of them is passed,
will be the most direct confirmation possible that the universe
underwent a brief period of acceleration ($d^{2}a/dt^{2}\!>\!0$)
followed by a long period of deceleration ($d^{2}a/dt^{2}\!<\!0$).
[Here $a(t)$ is the Robertson-Walker scale factor and $H(t)\equiv
d({\rm ln}\,a)/dt$ is the Hubble rate \cite{Weinberg}.]  Let us
briefly summarize the basic idea, with details postponed until the
next section.  First, note that $a(t)H(t)$ grows during acceleration,
and shrinks during deceleration.  According to the inflationary
scenario \cite{Weinberg}, any observable fourier mode of the
cosmological density field with comoving wavenumber $k_{\ast}$ reached
a moment during inflation known as ``horizon exit,'' at which the
ratio $a(t)H(t)/k_{\ast}$ was unity; then, during the period remaining
before the end of inflation, this ratio grew by $N_{{\rm
    bef}}(k_{\ast})$ e-folds; and finally, after inflation, the ratio
shrunk by $N_{{\rm aft}}$ e-folds, reaching the value of
$a_{0}H_{0}/k_{\ast}<1$ measured today.  Thus, in this picture, the
mode $k_{\ast}$ must satisfy the ``closure condition''
\begin{equation}
  \label{closure}
  {\rm ln}(a_{0}H_{0}/k_{\ast})=N_{{\rm bef}}(k_{\ast})-N_{{\rm aft}}.
\end{equation}

How can we test this equation?  If we know the temporal evolution of
$H$, from the moment that $k_{\ast}$ leaves the horizon until the end
of inflation, we have enough information to compute $N_{{\rm
    bef}}(k_{\ast})$ and $N_{{\rm aft}}$ and check that Eq.~(1) is
correct.  Unfortunately, cosmological observations will never give us
this much information; instead, assuming inflation is correct, they
will provide the first few terms in a Taylor approximation to $H$
around the moment that the ``WMAP wavenumber'' \cite{Komatsu:2008hk}
$k_{\ast}=0.002/{\rm Mpc}$ left the horizon.  Imagine we only have
enough observations to determine this Taylor approximation up to $j$th
order: this is our best guess for $H$ given the available data.  Under
the assumption that this guess remains valid all the way to the end of
inflation, we can check whether Eq.~(\ref{closure}) is true.  If it
is, then we not only have evidence for our guess, but for the idea of
inflation on which it is based: we have pulled ourselves up by our
bootstraps!  Note that $a_{0}H_{0}/k_{\ast}$, as measured today, is
exponentially sensitive to both $N_{{\rm bef}}(k_{\ast})$, which
depends on the expansion history \emph{during} inflation, and $N_{{\rm
    aft}}$, which depends on the expansion history \emph{after}
inflation.  Thus, if observations can be used to determine all three
quantities and if they are shown to satisfy the closure condition,
even an ardent skeptic would be hard-pressed to discount it as
coincidence; it would be strong evidence for inflation, and a tall
challenge for any competing theory.  And if there are any doubters,
the bootstrap method offers in some cases a series of follow-up checks
that can turn a convincing verification into an overwhelming one.

As discussed in the conclusion, the closure test also has the
advantage that it is experimentally easier to apply compared to other
proposed inflationary tests, such as the ``consistency relations"
\cite{consistency_gwaves, consistency_bispectrum}.  As for failing the
closure test, this does not mean inflation is ruled out, because it is
always possible to construct inflationary models that fit the data; as
precision improves, observers can continue to distinguish viable and
non-viable models.  However, in this sad circumstance, cosmological
observations will probably never yield the kind of convincing
confirmation of inflation discussed here.

\section{Key observables and equations}
\label{S:basic}

Before describing the bootstrap tests, let us introduce the key
observables, equations and parameters we will need.  From
$\Delta_{{\cal R}}^{2}(k)$ and $\Delta_{h}^{2}(k)$, the scalar and
tensor power spectra, one defines the tensor/scalar ratio
$r\!\equiv\!\Delta_{h}^{2}/ \Delta_{{\cal R}}^{2}$, the scalar index
$n_{s}\!-\!1\!\equiv\!d({\rm ln}\,\Delta_{{\cal R}}^{2})/d({\rm
  ln}\,k)$, its ``running'' $\alpha_{s}\!\equiv\!dn_{s}/ d({\rm
  ln}\,k)$, the ``running of the running''
$\beta_{s}\!\equiv\!d\alpha_{s}/d({\rm ln}\,k)$, and so on, all
measured at the wavenumber $k_{\ast}$ at which they are most precisely
determined.  We follow the standard WMAP definitions for all of these
observables \cite{Komatsu:2008hk}; and for concreteness fix
$k_{\ast}=0.002/{\rm Mpc}$ following WMAP \cite{Komatsu:2008hk}.

We focus first on the case where the Hubble expansion rate $H$ during
inflation is governed by a single order parameter that acts just like
a single inflaton scalar field, $\varphi$ \cite{Hubble_formalism,HSR}
\footnote{For our purposes, using $H(\varphi)$ has an advantage over
  the other common alternative $V(\varphi)$.  Namely,
  Eqs.~(\ref{end_eq},\ref{N_bef},\ref{N_aft}), which play an important
  role in our analysis, are exact in terms of $H(\varphi)$, whereas
  the corresponding equations, when expressed in terms of
  $V(\varphi)$, invoke the slow roll approximation near the end of
  inflation, where it is expected to break down.}  (Generalizations
will be discussed in the concluding section.)  For a clear derivation
and presentation of all of the equations in this section, see Liddle
{\it et al}\ \cite{Hubble_formalism}.  Subscripts ``$\ast$'' and
``end'' will be used to indicate that the corresponding quantity is to
be evaluated at the moment when $k_{\ast}$ exits the horizon, or at
the end of inflation, respectively.  Without loss of generality, we
can choose $\varphi_{\ast}=0$, and Taylor expand:
\begin{equation}
  \label{Taylor}
  H(\varphi)\!=\!H_{\ast}+H_{\ast}'\varphi+\frac{1}{2}H_{\ast}''
  \varphi^{2}+\frac{1}{6}H_{\ast}'''\varphi^{3}+\ldots 
\end{equation}
If, again without loss of generality, we take $d\varphi/dt>0$ (or,
equivalently, $H_{\ast}'<0$), and choose ``reduced Planck units'' with
$\hbar=c=8\pi G=1$, then the first few coefficients are
\begin{subequations}
  \label{Hi}
  \begin{eqnarray}
    \label{H0}
    H_{\ast}\!&=&\!\frac{\pi(\Delta_{{\cal R}}^{2})^{1/2}}{2}(2r)^{1/2}, \\
    \label{H1}
    H_{\ast}'\!&=&\!\frac{\pi(\Delta_{{\cal R}}^{2})^{1/2}}{8}(-r), \\
    \label{H2}
    H_{\ast}''\!&=&\!\frac{\pi(\Delta_{{\cal R}}^{2})^{1/2}}{32}(2r)^{1/2}
    [r+4(n_{s}-1)], \\
    \label{H3}
    H_{\ast}'''\!&=&\!\frac{\pi(\Delta_{{\cal R}}^{2})^{1/2}}{128}
    [64\alpha_{s}-3r^{2}-20r(n_{s}-1)].
  \end{eqnarray}
\end{subequations}
The end of inflation ($H=H_{end}$ and $\varphi=\varphi_{end}$) occurs
when $\ddot{a}=0$ or, equivalently,
\begin{equation}
  \label{end_eq}
  H_{end}=-\sqrt{2}\,H_{end}'\,.
\end{equation}
Finally, in the closure condition (\ref{closure}), we have
\begin{equation}
  \label{N_bef}
  N_{{\rm bef}}(k_{\ast})={\rm ln}(H_{end}/H_{\ast})\!-\!
  \frac{1}{2}\!\int_{\varphi_{\ast}}^{\varphi_{end}}\!d\varphi
  [H(\varphi)/H'(\varphi)]
\end{equation}
and
\begin{equation}
  \label{N_aft}
  N_{{\rm aft}}={\rm ln}[\Omega_{{\rm rad}}^{1/4}(H_{end}/H_{0})^{1/2}]
  +\Delta N,
\end{equation}
where $\Omega_{{\rm rad}}$ is the current ratio of the radiation
density to the critical density and
\begin{equation}
  \Delta N\equiv(1/12)\,\big[(1-3w_{re}^{})/(1+w_{re}^{})\big]\;{\rm ln}
  (\rho_{re}/\rho_{end})
\end{equation}
represents the uncertain physics of the epoch between the end of
inflation and the start of radiation domination: $w_{re}$ is the
effective equation-of-state during this epoch, and $\rho_{re}$ is the
energy density at the start of radiation domination.  We first
consider the case where $\Delta N \approx 0$, which corresponds to
``efficient'' reheating ($w_{re}\!\approx\!1/3$ or
$\rho_{re}\!\approx\!\rho_{end}$).  However, the uncertainty in
$\Delta N$, does not seriously interfere with the bootstrap test.  To
illustrate the point, we let $\Delta N$ be a free parameter, subject
only to the weak assumptions that $0\!\leq\!w_{re}\!\leq\!1/3$ and
$\rho_{\rm bbn} \!\leq\!\rho_{re}\!\leq\!\rho_{end}$, where $\rho_{\rm
  bbn}\!\approx\!(1~{\rm MeV})^{4}$ is roughly the energy density
during big bang nucleosynthesis (BBN); then $\Delta N\leq0$.  

\section{The bootstrap tests}
\label{S:bootstrap}

The bootstrap test uses precise observations at $k=k_{\ast}$ to obtain
progessively better estimates of $H(\varphi)$ and $H_{end}$, which
are, then, applied to determine if the closure condition is satisfied.
If we regard $\Delta_{{\cal R}}^{2}=(2.45\pm0.1)\times10^{-9}$
\cite{Komatsu:2008hk} as an already-measured quantity, then the Taylor
expansion (\ref{Taylor}, \ref{Hi}) of $H(\varphi)$ organizes the
remaining observables into an ordered list $\{r, n_{s}, \alpha_{s},
\beta_{s},\ldots\}$ in the sense that, if we imagine that we only know
the first $j$ items in this list, then we can only determine the
Taylor expansion up to $j$th order \footnote{Note that Taylor ordering
  $\{r,n_{s},\alpha_{s},\beta_{s},\ldots\}$ is not the order in which
  these quantities are detected in practice: $n_{s}$ has already been
  measured, but non-zero values for the remaining quantities
  (including $r$ and $\alpha_{s}$) have not yet been detected
  \cite{Komatsu:2008hk}.  Still, the Taylor ordering is what is
  relevant for conducting bootstrap tests: for example, until one
  detects the first observable $r$, none of the bootstrap relations
  may be confirmed.}.  This is the best guess for $H(\varphi)$ based
on the available data; using it, $H_{end}$ is computed from
Eq.~(\ref{end_eq}); $N_{{\rm bef}}(k_{\ast})$ and $N_{{\rm aft}}$ are
determined from Eqs.~(\ref{N_bef}, \ref{N_aft}); and finally the
closure condition (\ref{closure}) is checked.

The $j$th bootstrap test is satisfied if the first $j$ observables
satisfy the closure condition.  In practice, only the first three
observables $\{r,n_{s},\alpha_{s}\}$ can be detected or constrained
tightly enough to be relevant for confirming inflation.  Therefore,
the first three bootstrap tests are the relevant ones, for all
practical purposes: let us describe them and explain how they may be
confirmed and cross-checked with forthcoming observations.

{\bf First bootstrap test.}  To start, imagine we are only given the
first observable, $r$, so our best guess for $H(\varphi)$ is
$H_{\ast}+H_{\ast}'\varphi$.  We introduce this into
Eq.~(\ref{end_eq}) to obtain $\varphi_{end}=2^{1/2}[(16/r)^{1/2}-1]$,
and apply these expressions for $H(\varphi)$ and $\varphi_{end}$ to
Eqs.~(\ref{N_bef}, \ref{N_aft}) to obtain $N_{{\rm bef}}(k_{\ast})$
and $N_{{\rm aft}}$.  Then, the closure condition,
Eq.~(\ref{closure}), is satisfied if
\begin{equation}
  \label{1st_bootstrap}
  r(\Delta N)=8/[A+\Delta N+1/2]
\end{equation}
where $A\equiv{\rm ln}(a_{0}H_{0}/k_{\ast}) +\frac{1}{4}{\rm
  ln}(8\Omega_{{\rm rad}}\pi^{2}\Delta_{{\cal R}}^{2}/
H_{0}^{2})\approx61$.  This corresponds to $r=0.13$ if $\Delta N=0$,
and $0.13<r<0.17$ if the uncertainty in $\Delta N$ is included.

If observations pass this first bootstrap test, it will be a
remarkable success for the inflationary paradigm, and one that can be
checked: since true success should not be spoiled by the next
observable, $n_{s}$, we expect $n_{s}=1-r/4$ (so that
$H_{\ast}''\approx0$).  If this follow-up test is \emph{also}
successful, then it should not be spoiled by the next observable,
$\alpha_{s}$: thus we expect $\alpha_{s}=[3r^{2}+20r(n_{s}-1)]/64$ (so
that $H_{\ast}'''\approx0$).  If observations pass the first bootstrap
test (1), plus the two follow-up tests, it will be overwhelming
evidence for a period of inflationary expansion.

\begin{figure}
  \begin{center}
    \includegraphics[width=3.1in]{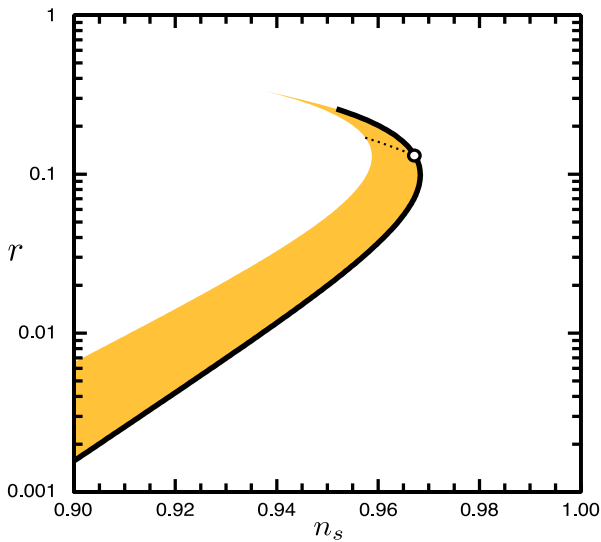}
  \end{center}
  \caption{First and second bootstrap tests: The first test is
    satisfied for $r$ and $n_s$ corresponding to the open white circle
    for efficient reheating; the circle becomes a short arc (dotted)
    if one includes the uncertainty in the ``reheating epoch''
    (parameterized by $\Delta N$).  The second test is passed if $r$
    and $n_s$ lie anywhere along the solid black curve or, allowing
    for the uncertainty in $\Delta N$, the black curve plus shaded
    region.  }
  \label{fig1}
\end{figure}

{\bf Second bootstrap test.}  If the first bootstrap test fails,
proceed to the second.  Given the measured values of $\{r,n_{s}\}$,
the best guess for $H(\varphi)$ is now
$H_{\ast}+H_{\ast}'\varphi+\frac{1}{2}H_{\ast}'' \varphi^{2}$.  We
introduce this expression into (\ref{end_eq}) and find the smallest
positive root: $\varphi_{end}(r,n_{s})$.  Next these formulae for
$H(\varphi)$ and $\varphi_{end}$ can be used in Eqs.~(\ref{N_bef},
\ref{N_aft}) to find $N_{{\rm bef}}(k_{\ast})$ and $N_{{\rm aft}}$.
Finally, substituting all of this into the closure condition
(\ref{closure}), we obtain a relation between $r$, $n_{s}$
corresponding to the solid black curve in Fig.~1; allowing for the
uncertainty in $\Delta N$ thickens the curve to the shaded region in
the figure.  Passing the second bootstrap test would be nearly as
remarkable as passing the first; and, here again, the success can be
verified by measuring the next observable, $\alpha_{s}$, and finding
that $\alpha_{s}=[3r^{2}+20r(n_{s}-1)]/64$ (so that $H_{\ast}'''
\approx 0$).

{\bf Third bootstrap test.}  If the first two bootstrap tests fail,
proceed to next order.  Use the first three observables
$\{r,n_{s},\alpha_{s}\}$ to determine the coefficients in $H(\varphi)
=H_{\ast}+H_{\ast}'\varphi+ \frac{1}{2}H_{\ast}''\varphi^{2}
+\frac{1}{6}H_{\ast}'''\varphi^{3}$.  Introduce this into
Eq.~(\ref{end_eq}) to find the smallest positive root
$\varphi_{end}(r,n_{s},\alpha_{s})$ and use the expressions for
$H(\varphi)$ and $\varphi_{end}$ in Eqs.~(\ref{N_bef}, \ref{N_aft}) to
compute $N_{{\rm bef}}(k_{\ast})$ and $N_{{\rm aft}}$.  Finally,
substitute these expressions into Eq.~(\ref{closure}) to check the
closure condition.  If $\Delta N=0$, the closure condition will be
satisfied for a 2-dimensional surface in the 3-dimensional space
parametrized by $\{r,n_{s},\alpha_{s}\}$: several surface contours are
shown in Fig.~2.  Allowing for the uncertainty in $\Delta N$ thickens
this surface into a ``thin slab'' (or, equivalently, each curve in
Fig.~2 extends \emph{downward} to form a strip).  In the limit of {\it
  extremely} small $r$, this slab has a simple analytic description:
$\varphi_{end}(r, n_{s}, \alpha_{s})$ is given by
$\varphi_{end}=(-2/\alpha_{s})^{1/2}r^{1/4}$, and
$r=r(n_{s},\alpha_{s},\Delta N)$ satisfies
\begin{equation}
  \label{3rd_bootstrap_small_r}
  r=8\,{\rm exp}\!\left[\frac{8\pi\!-\!16\,
      {\rm arctan}[(1\!-\!n_{s})/y]}{y}\!-\!4(A\!+\!\Delta N)\!\right]
\end{equation}
where $y\equiv[-4\alpha_{s}-(n_{s}-1)^{2}]^{1/2}$.  

We can make $\alpha_{s}$ as negative as possible (for fixed $n_{s}$)
by first letting $r$ be as small as possible [for illustration, let us
take the relatively weak assumption $\rho_{f}>(1~{\rm TeV})^{4}$ and
hence $r>8\times10^{-55}$], and then letting $\Delta N$ be as negative
as possible ($w_{re}=0$ and $\rho_{re}=\rho_{bbn}$).  In this way, we
find that, if $\{r,n_{s},\alpha_{s}\}$ pass the third bootstrap test,
then $\alpha_{s}$ has a lower bound \cite{Easther:2006tv}
$\alpha_{s}>\alpha_{s}^{{\rm min}}(n_{s})$, where $\alpha_{s}^{{\rm
    min}}(n_{s})$ varies smoothly from $\alpha_{s}^{{\rm
    min}}=-0.0094$ (for $n_{s}=0.9$) to $\alpha_{s}^{{\rm
    min}}=-0.0161$ (for $n_{s}=1$).

We have seen that, if the first bootstrap test is passed, then $n_{s}$
and $\alpha_{s}$ provide two cross-checks; and, similarly, if the
second bootstrap is passed, then $\alpha_{s}$ provides a single
cross-check.  But if the third bootstrap relation correctly predicts
that the $\{r,n_{s},\alpha_{s}\}$ lie in the slab described above,
there will be no analogous cross checks available, since we will have
used up our observables (see below for caveats). Nevertheless, passing
the third test is an impressive verification of the inflationary
principle.

\section{Discussion}

Observations will measure $r$ and $n_{s}$ to a precision of roughly
$\pm0.01$ in this decade, and perhaps $\pm0.001$ eventually. 
The first bootstrap test would give the most impressive proof of
inflation, since it makes the largest number of verifiable follow-up
predictions; the ranges $0.13<r<0.17$ and $n_{s}=1-r/4$ still agree
well with current observations \cite{Komatsu:2008hk}, but will either
be confirmed or ruled out within the next few years.  If the first
bootstrap test fails, the second may be passed for a wider range of
$r$; but since it relates $r$ to $n_{s}$, the allowed range of $r$ may
be restricted by constraining $n_{s}$.  For example, if $n_{s}>0.94$,
as suggested by WMAP5 \cite{Komatsu:2008hk}, then the second bootstrap
test only requires searching for $r>0.01$, and thus may also be
completed over the coming decade.  And then, if CMB polarization
experiments determine that $r<0.01$, all is not lost: the third
bootstrap test may still be passed, but only if $\alpha_{s}$ has a
substantially \emph{negative} value (see Fig.~2) -- {\it e.g.}\
negative enough to be detected by a proposed high-redshift galaxy
survey designed to measure $|\alpha_{s}|\sim0.001$ \cite{Komatsu}.
But if $\alpha_{s}$ is \emph{too} negative, all three bootstrap tests
fail: {\it e.g.} if $\alpha_{s}<-0.016$, then the tests fail for all
$n_{s}<1$, according to the discussion above.

\begin{figure}
  \begin{center}
    \includegraphics[width=3.1in]{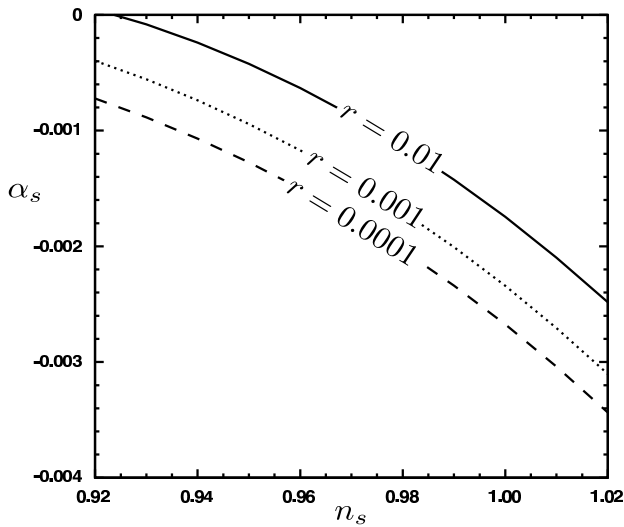}
  \end{center}
  \caption{The third bootstrap relation constrains $(r, \, n_s, \, 
    \alpha_s)$ to lie on a curve, as shown, assuming efficient
    reheating ($\Delta N =0$); including this uncertainty extends each
    curve downward (towards more easily observable values of
    $\alpha_s$) into a strip of finite thickness.  Note that these
    curves include values of $(r, \, n_s)$ that fail the first two
    bootstrap tests. }
  \label{fig2}
\end{figure}

Passing the bootstrap tests would be consistent with many of the most
appealing and commonly arising inflationary models, with the fewest
degrees of freedom, fewest parameters, and smoothest evolution
\cite{Boyle:2005ug}.  For example, in Fig.~1,
$V(\varphi)\!=\!m^{2}\varphi^{2}$ corresponds to the open circle (and
hence passes the first bootstrap test);
$V(\varphi)\!=\!\lambda\varphi^{4}$ corresponds to the upper endpoint
of the solid curve; the symmetry-breaking (Higgs) potential
$V(\varphi)\!=\!\Lambda^{4}[1-(\varphi/\mu)^{2}]^{2}$ corresponds to
the part of the solid curve \emph{below} the open circle; and the
pseudo-Nambu-Goldstone-boson (axion) potential
$V(\varphi)\!=\!\Lambda^{4}[1\pm{\rm cos}(\varphi/\mu)]$ lies within
the shaded region.  On the other hand, hybrid-like inflation models,
including some of the widely discussed proposals motivated by string
theory, would not pass the tests \cite{string_inflation}.

The bootstrap tests have the advantage that they can be performed with
forthcoming data.  Compare them with the well-known consistency
relations for single field inflation: (i) a hierarchy of relations
between the scalar and tensor power spectra \cite{consistency_gwaves};
and (ii) a hierarchy of relations between the primordial scalar
$N$-point functions \cite{consistency_bispectrum}.  Confirming any of
these relations requires measuring either (i) a non-zero value for the
tensor spectral index $n_{t}$, or (ii) a non-gaussian primordial
$N$-point function, both of which will be extremely difficult given
the single-field inflationary predictions.  Failure (detection of
large deviations from the consistency relations) is observationally
much easier than confirmation given the limitations of technology and
foregrounds.  By contrast, with the bootstrap tests, inflation can be
precisely tested and confirmed using accessible technology and
plausible foregounds. (The consistency relations might eventually
yield additional confirming tests.)

What if the bootstrap tests fail?  Of course, there will still be some
inflationary models and parameters that agree with the observations,
and some that do not.  But is there another way of confirming that
inflation itself took place?  Perhaps there is a generalized framework
for inflation that gives rise to a generalized set of bootstrap tests
which might still be passed?  Interestingly, the two most common
generalizations (allowing multiple order parameters \cite{hybrid,
  curvaton}, or replacing the canonical kinetic term
$X=\frac{1}{2}(\partial\varphi)^{2}$ by a general function of $X$
\cite{kinflate}) give rise to frameworks that are \emph{not} testable
in our bootstrap sense: the observables $\{\Delta_{{\cal R}}^{2},
r,n_{s},\alpha_{s},\ldots\}$ do not intrinsically point to a best
guess for both $N_{{\rm bef}}(k_{\ast})$ and $N_{{\rm aft}}$.  An
exception is the subclass of single-field $k$-inflation models
\cite{kinflate} which \emph{only} depend on $X$.  In this case, a
bootstrap test of (\ref{closure}) can be performed if one can
\emph{also} measure the tensor tilt $n_{t}$ (although, as mentioned
above, this is likely to be very difficult).  An interesting corollary
of our analyses is that, in all cases, a direct confirmation of
inflation relies on being able to detect cosmic gravitational waves
and measure accurately at least $r$.

We thank R.~Easther, J.~Frieman, W.~Kinney, E.~Komatsu and H.~Peiris
for discussions; and the Princeton Center for Theoretical Science for
its role in the development of this paper.  This work was supported by
a CIFAR Junior Fellowship (LB) and the US Department of Energy grant
DE-FG02-91ER40671 (PJS).


\begin{thebibliography}{99}

\bibitem{inflation} A.~H.~Guth, Phys.\ Rev.\ D {\bf 23}, 347 (1981);
  A.~D.~Linde, Phys.\ Lett.\ B {\bf 108}, 389 (1982); A.~Albrecht and
  P.~J.~Steinhardt, Phys.\ Rev.\ Lett.\ {\bf 48}, 1220 (1982).

\bibitem{Weinberg} S.~Weinberg, {\it Cosmology}, Oxford University
  Press (2008).

\bibitem{Komatsu:2008hk} E.~Komatsu {\it et al.}, arXiv:0803.0547.

\bibitem{consistency_gwaves} J.~E.~Lidsey {\it et al.}, Rev.\ Mod.\ 
  Phys.\ {\bf 69}, 373 (1997); M.~Cortes and A.~Liddle, Phys.\ Rev.\ D
  {\bf 73}, 083523 (2006).
  
\bibitem{consistency_bispectrum} J.~M.~Maldacena, JHEP {\bf 0305}, 013
  (2003); X.~Chen, M.-X.~Huang and G.~Shiu, Phys.\ Rev.\ D {\bf 74},
  121301 (2006).
  
\bibitem{Hubble_formalism} D.~S.~Salopek and J.~R.~Bond, Phys.\ Rev.\ 
  D {\bf 42}, 3936 (1990); A.~R.~Liddle, P.~Parsons and J.~D.~Barrow,
  Phys.\ Rev.\ D {\bf 50}, 7222 (1994).
  
\bibitem{HSR} W.~H.~Kinney, Phys.\ Rev.\ D {\bf 66}, 083508 (2002);
  A.~R.~Liddle, Phys.\ Rev.\ D {\bf 68}, 103504 (2003); H.~V.~Peiris
  and R.~Easther, JCAP {\bf 0607}, 002 (2006); JCAP {\bf 0610}, 017
  (2006); arXiv:0805.2154.
  
\bibitem{Easther:2006tv} R.~Easther and H.~Peiris, JCAP {\bf 0609},
  010 (2006).
  
\bibitem{Komatsu} E.~Komatsu, private communication.
  
\bibitem{string_inflation} A.~Linde, Phys.\ Rev.\ D {\bf 49}, 748
  (1994); 
  D.~Baumann and L.~McAllister, Phys.\ Rev.\ D {\bf 75}, 123508
  (2007); E.~Silverstein and A.~Westphal, arXiv:0803.3085.
  
\bibitem{Boyle:2005ug} L.~A.~Boyle, P.~J.~Steinhardt and N.~Turok,
  Phys.\ Rev.\ Lett.\ {\bf 96}, 111301 (2006)
  [arXiv:astro-ph/0507455].
  
\bibitem{hybrid} A.~D.~Linde, Phys.\ Rev.\ D {\bf 49}, 748 (1994).
  
\bibitem{curvaton} A.~Linde and V.~Mukhanov, Phys.\ Rev.\ D {\bf 56},
  535 (1997); D.~Lyth and D.~Wands, Phys.\ Lett.\ B {\bf 524}, 5
  (2002).
  
\bibitem{kinflate} C.~Armendariz-Picon, T.~Damour and V.~F.~Mukhanov,
  Phys.\ Lett.\ B {\bf 458}, 209 (1999) [arXiv:hep-th/9904075].

\end{thebibliography}
\end{document}